# Tuning Plasmonic Metasurfaces *via* Phase Change Material Substrates for Modulating Reactivity in Light-Driven Reactions

*Ning Lyu [a b], Anjalie Edirisooriya [b], Dawei Liu [c d], Zelio Fusco [b], Shenyou Zhao [e], Lan Fu [c], Fiona J. Beck [b*], Christin David [a f*]*

a. Institute of Solid State Theory and Optics, Friedrich-Schiller-Universität Jena, 07743 Jena, Germany

b. School of Engineering, Australian National University, ACT 2601, Australia

c. Australian Research Council Centre of Excellence for Transformative Meta-Optical Systems, Department of Electronic Materials Engineering, Research School of Physics, The Australian National University, Canberra ACT 2601, Australia

d. Institute of Applied Physics, Abbe Centre of Photonics, Friedrich Schiller University Jena, Albert-Einstein-Str. 15, Jena 07745, Germany

e. School of Electronic Engineering, Xi'an University of Posts and Telecommunications, Xi'an 710121, China

f. University of Applied Sciences Landshut, Am Lurzenhof 1, 84036 Landshut, Germany

* Corresponding authors: christin.david@uni-jena.de (C. David); fiona.beck@anu.edu.au (F.J. Beck);

**Abstract**

Phase change materials provide a powerful platform for dynamically modulating optical responses in nanophotonic systems. While plasmonic metasurfaces have been widely employed to enhance photocatalytic efficiency and promote particular light-driven reactions, active and dynamical control over reaction pathways within a single device remains challenging. Here, we report a phase-induced tunable metasurface that tailors photoexcited electron populations through mode hybridization, enabling selective control over the reactivity of light-driven chemical processes. By exploiting thermally induced refractive-index switching in a $Sb_2S_3$ cavity, the plasmonic resonance strength of Au nanodisks is actively tuned *via* cavity-plasmon hybridization. This reconfiguration modulates the product yield of methylene blue degradation by a factor of 2.4, suppressing to 0.45 in the crystalline phase and enhancing to 1.09 in the amorphous phase. Importantly, this reconfigurable platform enables dynamic control of the reaction yield using a single metasurface architecture under identical illumination conditions. Our approach establishes a dynamically programmable light-driven reaction platform capable of precisely manipulating reaction reactivity, offering new opportunities for selective photocatalysis in complex multibranch reaction systems.

**Introduction**

Light-driven reactions have recently attracted considerable interest across multiple disciplines because they convert absorbed photons into reactive pathways that accelerate the generation of valuable fuels[1–5], the synthesis of target organic products[6,7], and protein labeling and bioimaging[8,9]. Plasmonic metasurfaces, composed of metallic or metal-semiconductor hybrid nanostructures[10], support localized resonances and provide a platform for concentrating light within sub-wavelength regions where it can strongly interact with adsorbed molecular species[11]. In this context, metasurfaces offer the potential to catalyze specific reaction pathways and enhance reactivity toward desired products, addressing the demands of multibranched reaction systems such as $CO_2$ reduction[12,13]. Although thermodynamic and kinetic factors can govern catalytic outcomes[14], plasmonic metasurfaces enable a close-loop solution of driving the reaction transformation with sunlight and exhibit tailorable optical responses that enable dynamic coupling to specific reactions[15,16], thereby providing an opportunity for active modulation of catalytic reactivity.

Plasmonic metasurfaces drive light-induced reactions through localized surface plasmon resonances (LSPRs) and have emerged as a versatile platform for actively tuning photocatalytic activity[17,18]. Upon optical excitation at resonant wavelengths, collective oscillations of conduction electrons generate strongly enhanced near fields and energetic charge carriers. During plasmon decay, these carriers can populate the lowest unoccupied molecular orbital (LUMO) of adsorbed molecules through multiple pathways (Figure 1A), thereby initiating chemical transformations[1]. Specifically, near-field enhancement can promote direct intramolecular excitation from the highest occupied molecular orbital (HOMO) to the LUMO[17] (orange solid line in Figure 1A). Alternatively, plasmon dephasing produces hot electrons that inject into unoccupied molecular orbitals either directly via hybridized surface states (red solid line in Figure 1A) or indirectly through carrier diffusion processes (red dotted line) [12,19]. These mechanisms constitute non-thermal contributions of LSPR excitation, distinct from purely photothermal effects, yet equally capable of driving catalytic reactions[20]. Since catalytic activity strongly depends on the population of electrons in the LUMO[17,21], tailoring the optical response of plasmonic metasurfaces to spectrally align or misalign with molecular resonances provides a powerful handle for controlling reaction pathways. Such spectral engineering enables enhancement or suppression of charge transfer, thereby offering active modulation of light-driven reactivity steering production towards high-value chemicals.

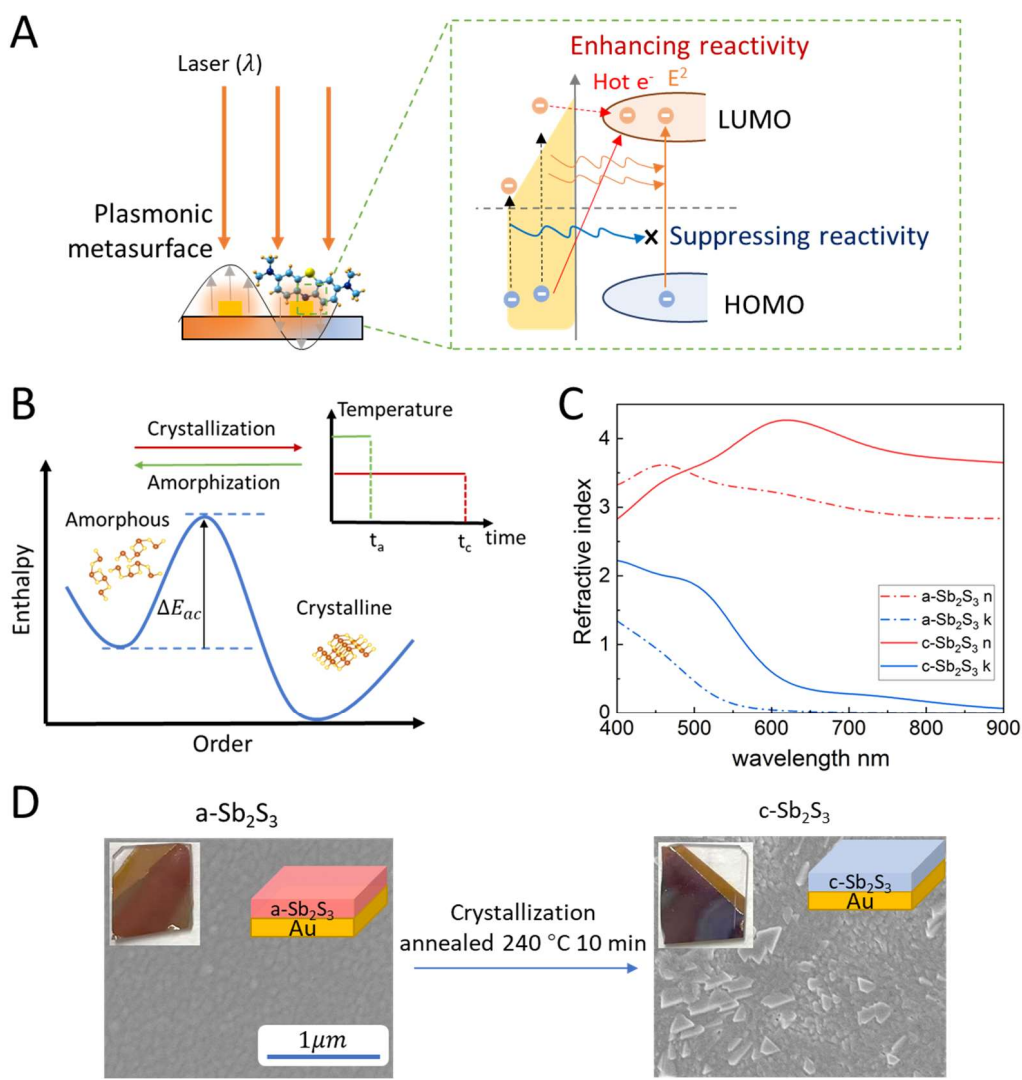

Figure 1: Tunable plasmonic resonance-driven reaction and phase transition of $Sb_2S_3$. A) Plasmon-mediated chemical reactions driven by the non-thermal effects of LSPRs. The resonance strength governs the population of photoexcited electrons transferred to the molecular LUMO, thereby modulating reaction reactivity. B) Temperature-time-dependent phase transformation of $Sb_2S_3$ between amorphous and crystalline states, illustrating the activation energy barrier overcome during thermally induced crystallization. C) Measured real (n) and imaginary (κ) components of the complex refractive index ($\tilde{n} = n + i\kappa$) for amorphous and crystalline $Sb_2S_3$. D) Surface configuration of $h_{Sb2S3}$ = 140 nm $Sb_2S_3$ thin films showing the transition from the amorphous state (left) to the crystalline state (right) after annealing at 240 °C for 10 min.

Plasmonic metasurfaces have emerged as powerful platforms for light-driven chemical transformations, enabling enhanced reactivity through the engineering of LSPRs that concentrate electromagnetic (EM) fields and generate energetic charge carriers. Early designs primarily focused on static resonance enhancement tailored to specific reactions[22–24]. In our previous work, we systematically investigated Au nanoparticles coupled to $TiO_2$ nanocavities with increasing cavity thickness, achieving up to a 102-fold modulation in product yield for methylene blue (MB) degradation through cavity-enhanced plasmonic coupling[25]. More recently, dynamically tunable plasmonic metasurfaces have enabled active control over reaction kinetics by modulating optical modes in situ. For example, Yuan *et al.* employed Ni nanoparticle-decorated $Si_3N_4$ nanocubes supporting quasi-bound states in the continuum (quasi-BICs), achieving polarization-controlled modulation of $H_2$ dissociation reactivity[26]. We further demonstrated polarization-dependent reactivity control using a single metasurface configuration based on elliptical Au-TiO$_2$ nanopillar arrays[27], in which the plasmonic resonance strength was modulated *via* incident light polarization. Despite these advances, the tunable range of plasmon-mediated reactivity remains constrained in compact metasurface architectures. And yet most nanophotonic architectures remain intrinsically static, with their optical response fixed by geometry after fabrication[28–30]. The plasmonic metasurfaces requires a compelling route toward expanding dynamic tunability by enabling dynamic modulation of coupled optical mode strength.

In parallel, phase change materials (PCMs) offer a powerful route to overcome this limitation by enabling strong, reversible modulation of optical properties without structural reconfiguration. Prototypical systems such as $Ge_2Sb_2Te_5$, $VO_2$ and $Sb_2S_3$ exhibit pronounced optical contrast across phase transitions, supporting non-volatile and on-demand functionality[30–33]. In particular, $Sb_2S_3$ combines a suitable bandgap[34], strong visible-light absorption and low toxicity[35] with sub-nanosecond switching and room-temperature phase stability[36], making it especially attractive for reconfigurable metasurfaces. A $Sb_2S_3$ cavity serves as an effective functional platform for metasurfaces, where the resonances of optically can be dynamically tuned through phase transitions of the underlying cavity.

As shown in Figure 1A, the enthalpy-structure diagram describes the temperature-time-dependent transformation between amorphous and crystalline phases of $Sb_2S_3$.[37] Owing to its relatively large bandgap, amorphous $Sb_2S_3$ (a-$Sb_2S_3$) exhibits low optical loss around wavelengths of 600 nm compared with other PCMs.[38,39] Crucially, the large refractive-index contrast ($|\Delta n| > 1.0$) between its phases enables substantial modulation of the optical response (Figure 1C).[29,32,40] Upon crystallization, bandgap narrowing induces a redshift in the extinction coefficient ($\kappa$), extending the cutoff wavelength to ~910 nm. Although the amorphous phase is thermodynamically less stable than the crystalline phase[38], crystallization proceeds upon overcoming an activation barrier via thermal annealing[40,41] (Figure 1D) or photothermal excitation[37,40]. When employed as the active layer of a Fabry-Pérot (F-P) cavity, simulated absorption (Figure 2D and E) reveals pronounced optical tunability[22,42]. Resonance redshifts are attributed to the increase in the real refractive index ($n$) upon crystallization, whereas the broadened wavelength absorption in the crystalline state arises from its reduced bandgap.

Here, we introduce a hybrid plasmonic metasurface incorporating a PCM cavity to actively manipulate photocatalytic reactivity through resonance hybridization. Thermal switching of $Sb_2S_3$ from amorphous to crystalline within the nanocavity modulates the F-P resonance and its coupling with the LSPR, thereby tailoring the population of photoexcited electrons at the molecular LUMO.

Using N-demethylation of MB as a model reaction, we experimentally demonstrate active enhancement and suppression of reactivity within a single metasurface configuration. In the amorphous phase, spectral alignment between the hybridized resonance and the molecular transition energy (corresponding to the HOMO-LUMO gap at 633 nm) selectively activates the demethylation reaction pathway. Upon crystallization, the resonance red-shifts and detunes from the molecular transition, leading to pronounced suppression of the reaction. This phase-dependent resonance engineering enables programmable control over particular reactions and product yields. Our work establishes a temperature- and time-programmable metasurface platform based on PCM-mediated resonance reconfiguration, providing a general strategy for actively manipulating plasmon-driven chemical reactivity.

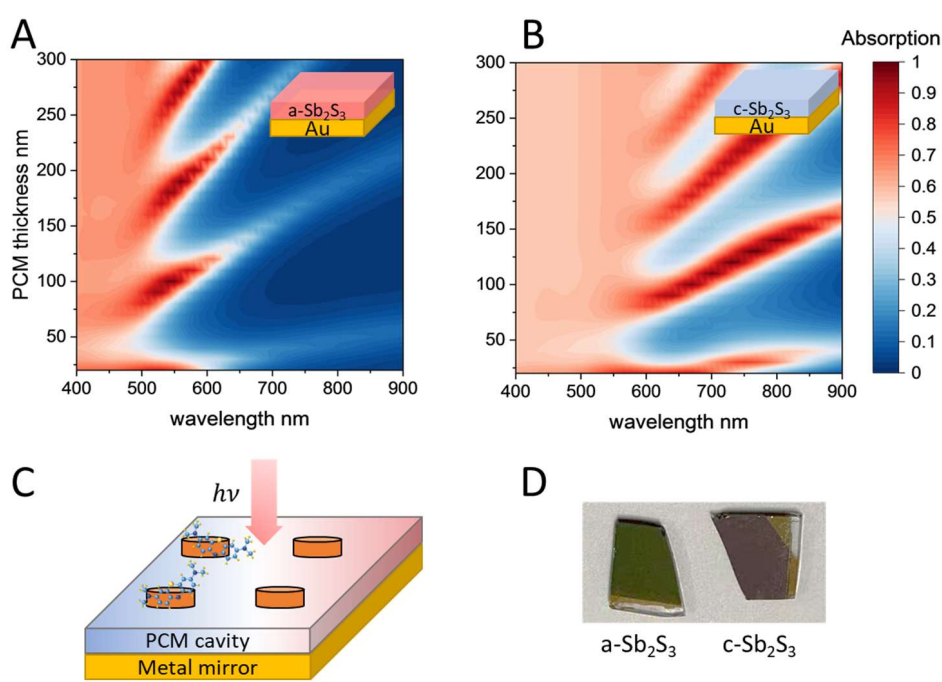

*Figure 2: Simulated absorption of $Sb_2S_3$ cavity and plasmonic metasurface configuration. A) Simulated absorption spectra of an amorphous $Sb_2S_3$ cavity on an Au mirror substrate with film thickness varying from 40 to 300 nm. B) Simulated absorption spectra of a crystalline $Sb_2S_3$ cavity on an Au mirror substrate with thickness ranging from 40 to 300 nm, showing constructive interference peak shifts arising from refractive-index variation upon phase transition. B) Schematic illustration of the Au nanodisk (ND)-$Sb_2S_3$ cavity metasurface configuration employed for light-driven catalysis. C) Images of Au ND-$Sb_2S_3$ cavity samples ($h_{Sb2S3}$ = 140 nm) in the amorphous (left) and crystalline (right) phases. Color difference reveals the shifting of optical properties.*

**Results**

**Tunable metasurface design and simulation**

The tunable metasurface is designed by integrating a phase change material (PCM), antimony sulfide ($Sb_2S_3$), as the cavity medium of a previously demonstrated hybrid plasmonic metasurface platform[25], shown schematically in Figure 2B. A $Sb_2S_3$ film is deposited onto a 200 nm-thick Au mirror layer supported on a glass substrate, forming a semi-open F-P cavity in which high reflectivity is sustained at the Au/$Sb_2S_3$ interface. An array of Au nanodisks (NDs) is subsequently fabricated atop the $Sb_2S_3$ layer, with a disk radius ($R_{Au\,ND}$) of 55 nm and a lattice periodicity ($a_{Au\,ND}$) of 150 nm. These geometrical parameters are selected to engineer the LSPR of the Au NDs to spectrally align with the excitation wavelength (633 nm) of the photodriven N-demethylation of MB. The wavelength of the F-P resonance is governed by the optical path length of the cavity, determined by the refractive index and thickness of the $Sb_2S_3$ layer. The $Sb_2S_3$ thickness is precisely tailored to position the Au ND-$Sb_2S_3$ cavity resonance at the desired wavelength. When the F-P resonance spectrally and spatially overlaps with the LSPR of the Au NDs, strong mode coupling can occur[43], resulting in hybridized resonances that enhance the plasmonic field intensity.

The Au ND-$Sb_2S_3$ cavity samples in their amorphous and crystalline phases are shown in Figure 2C, exhibiting a pronounced color contrast arising from the refractive index change accompanying the phase transformation. To further verify the structural transition, standalone amorphous and crystalline $Sb_2S_3$ (a-/c-$Sb_2S_3$) thin films were characterized by Raman spectroscopy before and after thermal annealing. As presented in the Supplementary Information (SI Figure S1), the amorphous film prior to annealing displays broad Raman peaks, corresponding to Sb-S and S=S vibrational modes, respectively.[40] Following annealing, the spectrum evolves into sharper and well-defined peaks characteristic of c-$Sb_2S_3$, confirming the formation of the orthorhombic crystalline phase[41], verifying the thermally induced phase transition of $Sb_2S_3$ within the cavity structure.

The optical tunability of the Au ND-$Sb_2S_3$ cavity in both phases is investigated using finite-element simulations (COMSOL Multiphysics). Total absorption spectra are calculated as a function of cavity thickness while maintaining fixed Au ND array parameters ($R_{Au\,ND}$ = 55 nm and $a_{Au\,ND}$ = 150 nm), as shown in Figure 3A and B. Compared to simulations of the PCM film alone in Figure 1D and E, incorporation of the Au ND array substantially broadens the absorption range due to excitation of LSPR, extending the spectral response to approximately 900 nm. The LSPR of the Au ND array is centered near 660 nm, closely matching the excitation wavelength employed for the photocatalytic reaction. This resonance position is predominantly governed by the ND geometry.

In the amorphous configuration, the LSPR of the Au ND hybridizes with the F–P cavity resonance at specific $Sb_2S_3$ thicknesses, notably $h_{Sb2S3}$ = 40 nm and 140 nm, corresponding to the first- and second-order F-P modes near the target wavelength. At these thicknesses, pronounced spectral splitting is observed, indicating strong coupling between the plasmonic and cavity modes. This hybridization occurs under constructive interference conditions within the cavity, leading to enhanced electromagnetic confinement and increased absorption at the target wavelength. As shown in Figure 3C and D, the absorption reaches ~80% and ~90% at 633 nm for $h_{Sb2S3}$ = 40 nm and 140 nm, respectively. The coupling strength ($\Omega$), extracted from the frequency separation of the split modes, is 56.9 GHz for the 40 nm cavity and 40.1 GHz for the 140 nm cavity, confirming the formation of hybridized plasmon-cavity states[44,45].

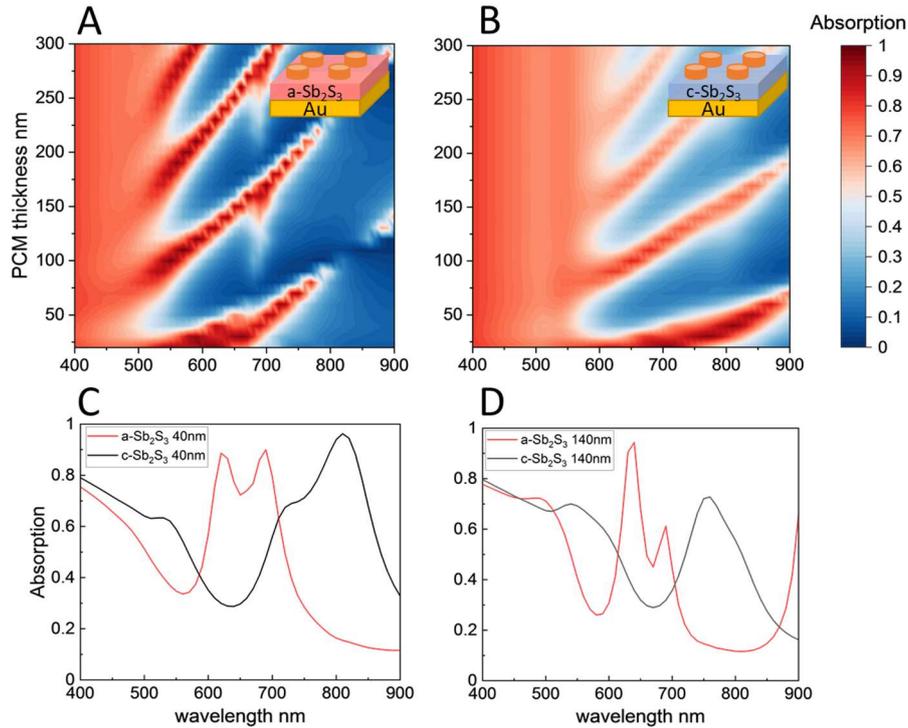

*Figure 3: Simulated total absorption spectra of the Au ND-$Sb_2S_3$ cavity in the A) amorphous and B) crystalline phases, calculated for cavity thicknesses ranging from 40 to 300 nm. At the target wavelength of 633 nm, the plasmonic resonance strongly hybridizes with the F–P cavity modes in the amorphous configuration at C) $h_{Sb2S3}$ = 40 nm and D) 140 nm, resulting in enhanced absorption due to constructive interference and mode coupling. In contrast, in the crystalline phase, the cavity resonance red-shifts and supports destructive interference near 633 nm, weakening plasmon–cavity hybridization and leading to a reduction in absorption.*

In contrast, upon crystallization of $Sb_2S_3$, the increased refractive index shifts the cavity resonance toward longer wavelengths, while the higher absorption coefficient introduces additional optical losses. Consequently, the spectral splitting is substantially reduced, reflecting weakened mode hybridization. For the metasurface with $h_{Sb2S3}$ = 40 nm, the resonance peak red-shifts to ~810 nm in the crystalline phase, and the absorption at 633 nm decreases from ~80% to 29%. Similarly, for $h_{Sb2S3}$ = 140 nm, the resonance shifts to ~760 nm, reducing absorption at 633 nm from ~90% to 41%.

This comparison highlights a pronounced phase-dependent modulation of optical absorption within an identical structural geometry. The PCM-integrated metasurface therefore functions as an optical switch, toggling absorption at the target wavelength between "on" (amorphous) and "off" (crystalline) states *via* controlled phase transformation. Consistent trends are observed in simulations of non-thermal LSPR-driven effects (Figure S2), where variations in hot-electron generation and near-field enhancement correlate directly with the calculated absorption behavior. Because phase switching in $Sb_2S_3$ can occur on sub-nanosecond timescales in confined geometries, this platform could be integrated with localized thermal control schemes to enable real-time modulation of plasmonic resonance and catalytic activity.

**Sample preparation and optical property characterization**

The Au ND-$Sb_2S_3$ cavity metasurfaces are fabricated using an etch-free process based on the geometrical parameters optimized through numerical simulations (see Methods and SI Figure S4). The $Sb_2S_3$ cavity layer is deposited by thermal evaporation to precisely controlled thicknesses of 40, 140, and 160 nm. $Sb_2S_3$ cavities are in the amorphous phase. The cavity thickness is tuned by adjusting the evaporation duration. To obtain crystalline-phase samples, the amorphous $Sb_2S_3$ films are annealed at 240 °C for 10 min, ensuring complete phase transformation while maintaining film uniformity and reducing pyramidal crystalline domains[41] (see SI Figure S6 and S7). Subsequently, the Au ND arrays are fabricated atop the $Sb_2S_3$ layer with electron-beam lithography (EBL). The morphology of the resulting metasurfaces is characterized by scanning electron microscopy (SEM), as shown in Figure 4A and B. The measured Au ND radius is approximately 59 nm, with an array periodicity of 150 nm, in good agreement with the intended design.

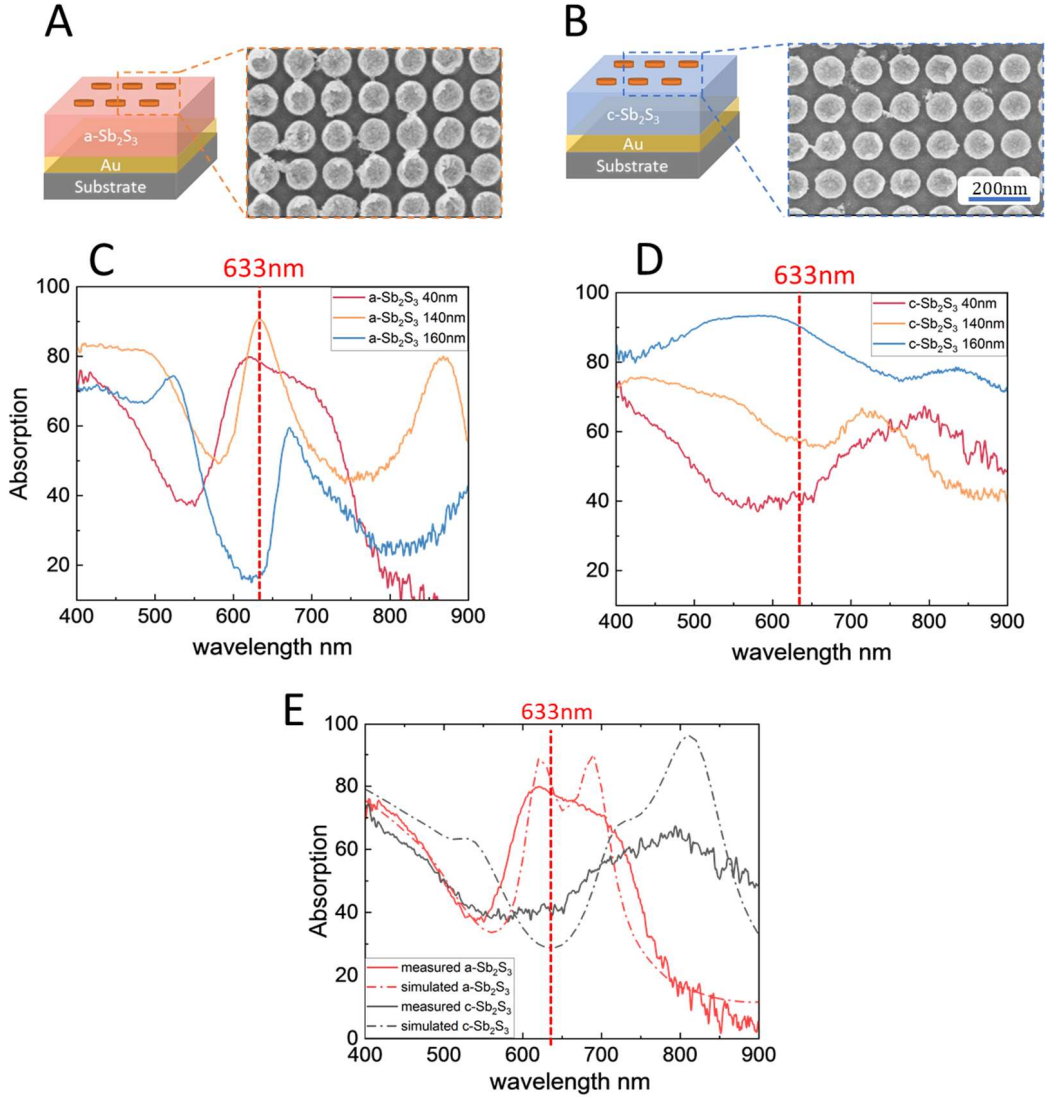

*Figure 4: Schematic of the Au ND-Sb₂S₃ cavity metasurface and SEM images of the Au ND arrays for the A) amorphous and B) crystalline phases. Experimentally measured absorption spectra of Au ND-Sb₂S₃ cavity samples with Sb₂S₃ thicknesses of $h_{Sb2S3}$ = 40, 140, and 160 nm in the C) amorphous and D) crystalline phases. E) Comparison between measured and simulated absorption spectra of the 40 nm thick Au ND-Sb₂S₃ cavity in the amorphous and crystalline states, highlighting phase-dependent tunability of absorption at the target wavelength of 633 nm.*

The optical properties of the amorphous and crystalline metasurfaces are characterized using a spectrophotometer. The phase-switching-induced tunability of the Au ND-Sb₂S₃ cavity is presented in Figure 4C and D for different cavity thicknesses. For the sample with $h_{Sb2S3}$ = 140 nm, the plasmonic resonance of the Au NDs couples to the second-order F–P cavity mode (m = 2). In the amorphous state (Figure 4C), the metasurface exhibits a pronounced absorption peak of 95% at the target wavelength of 633 nm. Upon crystallization (Figure 4D), the absorption decreases to 59%, demonstrating phase-dependent optical modulation. As the cavity thickness increases, the resonance peak progressively red-shifts. For example, in the amorphous sample with $h_{Sb2S3}$ = 160 nm, the resonance shifts to 665 nm, consistent with the expected scaling of the F-P cavity mode with optical path length. This shift repositions the cavity resonance relative to the target reaction wavelength.

Although clear tunability is observed for the $h_{Sb2S3}$ = 140 nm sample, the experimentally measured modulation range (a difference of ~36%) is smaller than predicted by simulations. In particular, the crystalline-phase sample retains relatively high absorption at 633 nm despite the expected detuning of plasmon-cavity coupling. Ellipsometer measurements (Table S1) indicate that the overall cavity thickness remains nearly unchanged after annealing for all samples. However, the surface roughness increases slightly in the crystalline phase, likely due to enhanced surface tension associated with structural reorganization during crystallization. For thicker $Sb_2S_3$ layers, the accumulated internal stress may further increase surface roughness, thereby degrading cavity quality and weakening the contrast between constructive and destructive interference conditions. To improve modulation efficiency, the cavity design was therefore optimized to target the first-order F-P resonance (m = 1), corresponding to $h_{Sb2S3}$ = 40 nm, where reduced thickness decreased roughness-induced losses and enhances phase-dependent tunability.

Figure 4E compares the simulated and measured absorption spectra of the Au ND-$Sb_2S_3$ cavity with $h_{Sb2S3}$ = 40 nm in the amorphous and crystalline phases, highlighting the enhanced optical tunability achieved through phase transformation. At the target wavelength for MB photocatalysis (633 nm), the absorption is modulated from 78% in the amorphous state to 39% in the crystalline state, corresponding to an effective tunability in plasmonic resonance strengths. The experimental spectra follow the same overall trend as the simulations, confirming phase-dependent resonance reconfiguration. Minor discrepancies are observed in the peak splitting, which is less pronounced experimentally than predicted numerically. This difference likely arises from fabrication-induced deviations that broaden the plasmonic and cavity resonance peaks. Instead of clearly resolved dual peaks, the amorphous sample exhibits a broadened absorption band, indicative of strong plasmon–cavity coupling near the F-P resonance maximum and consistent with significant resonance enhancement through hybridization.

Notably, an additional resonance feature appears near 800 nm in the crystalline state, where the absorption reaches 68%, compared to 17% in the amorphous phase. This phase-dependent modulation (~51%) provides an alternative operational wavelength window for photocatalytic control. Such substantial spectral contrast enables selective activation of reaction pathways aligned with different excitation wavelengths, suggesting strong potential for programmable reactivity switching. The $h_{Sb2S3}$ = 40 nm Au ND-$Sb_2S_3$ cavity represents an optimized configuration that supports large absorption modulation across multiple spectral regions, offering a versatile platform for dynamically controlling diverse plasmon-driven photochemical reactions.

**Light-driven reaction characterization**

To evaluate the catalytic performance of the Au ND-$Sb_2S_3$ cavity, surface-enhanced Raman spectroscopy (SERS) is employed to monitor the reaction in real time. The MB degradation reaction is driven by plasmonic excitation, which promotes photochemical N-demethylation to generate thionine and other intermediates with distinct vibrational signatures[46]. The characteristic product peak at ~480 cm$^{-1}$ serves as a mark allowing a quantitative analysis in indicative product yield[21], following the methodology described in previous work[17,21] and in the Methods section. Temporal Raman spectra are recorded to track the evolution of reaction products over time by monitoring characteristic peaks associated with N-demethylation of MB. Reaction schemes and additional experimental details are provided in Figure S9 and the Methods section.

As a phase change material, $Sb_2S_3$ can undergo thermally induced transitions, which may be triggered unintentionally by photothermal effects under laser illumination during SERS measurements, particularly in the amorphous state. To minimize such effects, all measurements are conducted under a controlled temperature environment held at 15 °C using a Linkam thermal stage. Maintaining a consistent temperature is critical, as reaction kinetics are strongly temperature dependent. Additionally, a low excitation laser power of 1.4 mW is concentrated on samples with a 20× objective lens to prevent unintended phase switching, and the microscope imaging system ensured the sample remained in a stable phase throughout the measurement.

Raman spectra are recorded with a 1 s interval over a 100 s period for both amorphous and crystalline Au ND-$Sb_2S_3$ cavities, then normalized with benzene ring peak at ~1622 cm$^{-1}$. As shown with the product peak at 480 cm$^{-1}$ in Figure 5A, the crystalline-phase sample exhibits only a modest increase in normalized peak intensity to 0.13 after 100s, reflecting suppressed catalytic activity. In contrast, the amorphous Au ND-$Sb_2S_3$ cavity demonstrates substantially enhanced reactivity, with the product peak rising to a normalized value of 0.24, consistent with stronger plasmonic resonance and more efficient charge-carrier generation driving the photodegradation.

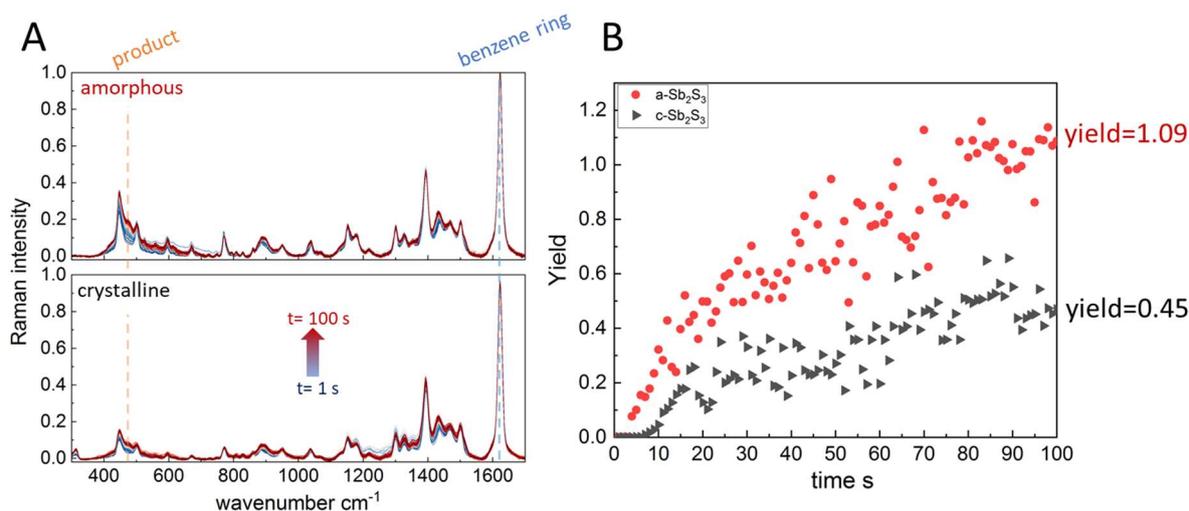

*Figure 5: A) SERS spectra of amorphous (top) and crystalline (bottom) Au ND-Sb$_2$S$_3$ cavity samples measured at 15°C in 100s duration. The spectrum collected in 1s timestep. B) Yield calculated by the integrated product peak at 480 cm$^{-1}$, which reflects the generation rate of N-demethylation derivates. At t = 100 s, the yield reaches 1.09 on amorphous Au ND-Sb2S3 cavity, compared with only 0.45 for the crystalline sample.*

Figure 5B shows the time evolution of indicative product yield for both amorphous and crystalline Au ND-Sb$_2$S$_3$ cavities. A noticeable increase in yield appears around 5 s for both phases, with the amorphous sample exhibiting a more significant initial slope. The reaction proceeds rapidly during the first 20 s, followed by a more gradual increase over the remainder of the 100 s measurement period. At the final timestep, the crystalline cavity reaches an indicative yield of approximately 0.45, whereas the amorphous cavity achieves a substantially higher indicative yield of 1.09, driven by the enhanced plasmonic resonance in the amorphous state. Repeatability measurements are presented in Figure S11. These results indicate that phase switching results in a 2.4-fold enhancement in reaction yield under identical experimental conditions.

The strength of the effect relative to the overall reactivity can be quantified by defining the switching amplitude[47] $A_s$ as:

$$A_s = \frac{\text{Yield}_{\text{amorphous}} - \text{Yield}_{\text{crystalline}}}{(\text{Yield}_{\text{amorphous}} + \text{Yield}_{\text{crystalline}})/2}$$

where $\text{Yield}_{\text{amorphous}}$ and $\text{Yield}_{\text{crystalline}}$ are the indicative product yields at the final measurement time for the two phases. For the $h_{\text{Sb2S3}}$ = 40 nm Au ND-Sb$_2$S$_3$ cavity, this $A_s$ is 0.83, highlighting the strong relative switching of reaction reactivity achieved through phase switching.

The modulation of reaction reactivity has been clearly demonstrated using the phase-transformation strategy in the Au ND-Sb$_2$S$_3$ cavity. The strength of the plasmonic resonance is effectively tuned through coupling with the F-P cavity mode, which in turn governs the population of photoexcited electrons in the reactant's LUMO. Since the resonance is wavelength-dependent, the metasurface selectively drives MB degradation along the N-demethylation pathway[21,27,46], which is quantitatively analyzed via SERS spectra. Phase switching of the $Sb_2S_3$ cavity tailors the plasmon-excited electron population through mode hybridization, allowing enhancement of reactivity in the amorphous phase and suppression in the crystalline phase. Notably, this modulation is achieved within a single metasurface configuration under the identical illumination condition which bordering the application environments.

By coupling plasmonic resonances with the PCM cavity, the absorption spectrum can be tuned over a broad range around the target excitation wavelengths of 633 nm and 785 nm, enabling effective modulation of the plasmonic resonance strength for reaction control[48]. The Au ND-$Sb_2S_3$ cavity exhibits tunable optical properties across multiple spectral regions, providing catalytic potential for diverse light-driven reactions. It can be applied to reactions whose molecular resonances align with additional cavity or plasmonic resonance peaks. For instance, the design supports significant absorption at 532 nm in both material phases, enabling an additional aligned photocatalytic reaction such as of Rhodamine 6G (R6G) degradation (Figure S15)[49]. In the crystalline phase, the metasurface promotes reactions near 800 nm, enhancing product yield. During phase switching, a wide absorption tunability (~51%) at this wavelength allows further optimization of reaction control.

The catalytic performance is compared with previous designs in Table 1. As a design with single metasurface configuration, the Au ND-$Sb_2S_3$ cavity supports a more effective switching amplitude compared with elliptical Au-$TiO_2$ nanopillar array, which also behave as a similar value with the quasi-BIC-Ni antenna under stronger light power and longer durations. The reason of this phenomenon can be contributed to the strong coupling of plasmonic resonance with both constructive peak and destructive node of cavity resonance to actively tailor its optical response. Similar trend can be observed in Au nanoparticle coupled cavities as well, which has the largest switching amplitude in 1.96 among all these approaches. Moreover, the Au ND-$Sb_2S_3$ cavity offer a dynamic strategy to modulating the reactivity, different with the all other designs relying on the changing or asymmetry in geometries. Here, the illumination conditions are kept consistent throughout the measurement, and the cavity phase are introduced into the system to controlling the reactivity with thermal-induced approaches, which has the potential to further extended to high-power laser-induced and electrical Jour heat induced material phase switching. However, the limited thermal stability of the amorphous Au ND-$Sb_2S_3$ cavity imposes constraints on environmental temperature and effective laser power in measurements[37,40], which limits the acceleration of photocatalytic reactions. While previous work from our group demonstrated the dominant contribution of non-thermal LSPR effects in MB degradation, environmental temperature still influences reaction kinetics by affecting the thermal energy of reactant molecules[17,50]. But the large switching amplitude has demonstrate an effective controlling in reaction reactivity, which hold the potential in achieving selectivity for target reaction. These results demonstrate the potential of the Au

ND-Sb$_2$S$_3$ cavity as a versatile platform for dynamical, and considerable reactivity modulating of light-driven reactions.

The catalytic performance is compared with previous designs in Table 1. As a single metasurface configuration, the Au ND-Sb$_2$S$_3$ cavity exhibits a more pronounced switching amplitude than the elliptical Au-TiO$_2$ nanopillar array, while achieving comparable performance to the quasi-BIC Ni antenna, despite the latter requiring higher illumination power and longer exposure times. This behavior can be attributed to strong coupling between the plasmonic resonance and both the constructive peak and destructive node of the cavity resonance, enabling active tailoring of the optical response. A similar trend is observed in Au nanoparticle coupled cavities, which display the largest switching amplitude (1.96) among the reported approaches. Moreover, the Au ND-Sb$_2$S$_3$ cavity provides a dynamic strategy for modulating reactivity that differs from previous designs relying on geometric variation or structural asymmetry. Here, the illumination conditions are maintained throughout the measurements, and phase changes within the cavity are introduced to control reactivity *via* thermal-induced transitions. This approach could be further extended to high-power laser-induced switching[40] or electrically driven Joule heating phase transitions[38]. However, the limited thermal stability of the amorphous Au ND-Sb$_2$S$_3$ cavity imposes constraints on the environmental temperature and effective laser power during measurements, which restricts the achievable acceleration of photocatalytic reactions. Although prior work from our group demonstrated the dominant contribution of non-thermal localized surface plasmon resonance effects in MB degradation, environmental temperature still influences reaction kinetics by modifying the thermal energy of reactant molecules. Nevertheless, the large switching amplitude demonstrated here provides effective control over reaction reactivity and holds promise for achieving selectivity toward target pathways. Collectively, these results highlight the Au ND-Sb$_2$S$_3$ cavity as a versatile platform for dynamic and substantial modulation of light-driven catalytic reactions.

Table 1. Features of the tunable metasurface strategies for light driven reaction control.

| Structure | Sample size mm$^2$ | Switching mechanism | Duration s | Light/ laser power mW | $A_s$ | Ref. |
|---|---|---|---|---|---|---|
| parallel zigzag Au nanoantenna arrays | - | optical chirality of incident light | - | - | 1.52 (simulated) | 47 |
| quasi-BIC-Ni antenna | $1 \times 1$ | polarization of incident light | 600 | 500 | 1.15 | 26 |
| Au nanoparticle coupled cavities | $\sim 10 \times 10$ | thickness of cavity | 10 | 0.68 | 1.96 | 25 |
| elliptical Au-TiO$_2$ nanopillar array | $0.1 \times 0.1$ | polarization of incident light | 10 | 6.6 | 0.72 | 27 |
| Au ND-Sb$_2$S$_3$ cavity | 0.1 × 0.1 | material phase change | 100 | 1.4 | 0.83 | This work |

Tailoring the morphology of the metasurface enables precise adjustment of optical properties to meet the requirements of more complex photochemical systems, such as CO$_2$ reduction[12,51]. This platform could provide the ability to selectively modulate specific reaction pathways, offering a route toward achieving product selectivity in multibranched light-driven catalytic systems. And since phase switching is reversible by laser-induce switching[38] this means that reactivity of the system can be dynamically modulated which can participating the dynamic evolution of real active sites[52].

**Discussion**

We have developed a switchable phase change metasurface, the Au ND-Sb$_2$S$_3$ cavity system, providing a predictable, dynamic, thermal-induced platform for controlling the plasmonic resonance strength of Au ND arrays *via* phase transitions of the cavity material. This tunability arises from tailored plasmonic resonance enhancement through hybridization with cavity modes, facilitated by refractive index changes during transitions between the amorphous and crystalline phases. Light-driven reaction reactivity can be selectively enhanced or suppressed depending on the Sb$_2$S$_3$ phase, which modulates resonance strength at wavelengths matched to the molecular HOMO-LUMO gap of the model reaction, the photodegradation of MB. This approach allows modulation of the indicative product yield from 0.45 to 1.09 with a considerable switching amplitude of 0.83 under identical illumination conditions on a single metasurface configuration. The Au ND-Sb$_2$S$_3$ cavity metasurface therefore offers a versatile strategy for actively tuning reaction yields and achieving selective catalysis across diverse photochemical systems.

**Method**

**FEM simulations.**

The Au NP-Sb$_2$S$_3$ cavity system was modeled using COMSOL Multiphysics, where the near-field and far-field optical properties were simulated using the finite element method (FEM). The optical characteristics of the samples were calculated in two separate phases corresponding to the amorphous and crystalline of Sb$_2$S$_3$. The refractive index of Sb$_2$S$_3$ used in the model was obtained from measurements of a 160 nm thin film deposited on an Au substrate, characterized by spectroscopic ellipsometry (JA Woollam M-2000D) over the wavelength range of 200 - 1700 nm, as shown in Figure 1B.

a periodic array of Au NDs was simulated by applying to Floquet periodic boundary conditions to the four lateral sides of the unit cell. The unit cell geometry consisted of a semi-infinite glass substrate, a 200 nm Au mirror layer, a Sb$_2$S$_3$ cavity layer, Au nanodisks, and an air layer. Incident light was normally coupled to the metasurface and introduced through a port on the top surface of the air layer. The total absorption was calculated using $A = 1 - T - R$, where T represents the total transmittance and R corresponds to the total reflectance. The near-field properties were analyzed through the normal component of the electric field. In addition, the cavity thickness and nanodisk parameters were swept to investigate their influence on the optical response.

**Sample preparation.**

Glass slides were used as the substrates for the samples. A 200 nm Au mirror layer was deposited by electron-beam evaporation (Temescal BJD-2000) under a vacuum pressure of $1 \times 10^{-5}$ Torr. Subsequently, Sb$_2$S$_3$ phase change material films were deposited onto the Au mirror by thermal evaporation with thicknesses of 40, 140, and 160 nm. The deposited Sb$_2$S$_3$ films were in the amorphous phase. Crystallization was achieved by annealing the samples on a hotplate at 240 °C for 10 minutes. Nanodisk patterns were then defined by electron-beam lithography (EBL) with exposure doses ranging from 675 to 725 $\mu C/cm^2$. A bilayer PMMA resist was used, consisting of PMMA 495 A2 and PMMA 950 A4. After pattern development, a 40 nm Au layer was deposited by electron-beam evaporation, followed by lift-off in acetone for more than 4 hours to remove the photoresist. This process resulted in the formation of the Au nanodisk array.

To facilitate molecular attachment, the fabricated samples were immersed overnight in a 5 ppm methylene blue solution.

**Surface Enhanced Raman Spectroscopy of MB degradation.**

Raman spectra of the samples in both amorphous and crystalline phases were measured using a Renishaw inVia Reflex Raman spectrometer. A 633 nm excitation laser was employed for the measurements. An Olympus LMPLFLN 20× objective lens was used for sample positioning and monitoring. A 1200 lines/mm grating was selected, which determines the spectral resolution of the system. During Raman measurements on the Au NP-Sb$_2$S$_3$ cavity structures, Raman signals were collected over a 100 s period with a temporal interval of 1 s between successive spectra to enable time-resolved analysis. The spectral baseline was removed using adaptive baseline correction methods to minimize the influence of background signals. For tracking of reaction evolution, the Raman spectra were normalized to the peak at ~1622 cm$^{-1}$, which corresponds to the C-C stretching vibration of the benzene ring.

In surface-enhanced Raman scattering measurements, the vibrational fingerprints revealed by Raman spectroscopy are unique to individual molecules and are therefore widely used for chemical identification. In particular, in situ real-time Raman measurements enable monitoring of the temporal evolution of chemical reactions. MB is a commonly used model molecule in SERS studies, as its spectroscopic characteristics are well understood. This makes MB an ideal probe molecule for evaluating the catalytic performance of the metasurface design investigated in this work. The products of N-demethylation reaction are monitored with the Raman peak at ~480 cm$^{-1}$. To quantitative analyze the reactivity, this product peaks were fitted with the deconvolution method based on a Gaussian-Lorentzian blend model. And the indicated yields were calculated by the integration of the fitted curves for revealing the time evolution of product generation.

## ASSOCIATED CONTENT

**Supporting Information**

The supporting information is shown at *Tuning plasmonic metasurfaces via phase change material substrates for modulating reactivity in light-driven reactions_SI.pdf*.

**Author Contributions**

The manuscript was written through contributions of all authors. All authors have given approval to the final version of the manuscript.

**Funding Sources**

The project is funded by International Research Training Group IRTG 2675 (GEPRIS 437527638).

## ACKNOWLEDGMENT

This work is supported by the German Research Foundation DFG (Deutsche Forschungsgemeinschaft) through funding of the International Research Training Group IRTG 2675 (GEPRIS 437527638). The authors acknowledge access to NCRIS facilities (ANFF-ACT Node) at the Australian National University. The authors acknowledge Mr. S. Debbarma for supporting with the Sb$_2$S$_3$ evaporation.